\RequirePackage{fix-cm}
\documentclass[twocolumn]{svjour3}
\smartqed
\usepackage{graphicx}
\usepackage{url}
\usepackage{color}
\journalname{Brazilian Journal of Physics}
\begin{document}

\title{Stochastic Background of Gravitational Waves Generated by Compact Binary Systems}

\titlerunning{ }

\author{Edgard F. D. Evangelista \and Jos\'{e} C. N. de Araujo}

\authorrunning{ }

\institute{Edgard F. D. Evangelista \at
              Instituto Nacional de Pesquisas Espaciais -- Divis\~{a}o de Astrof\'{i}sica, Av. dos Astronautas 1758, S\~{a}o Jos\'{e} dos Campos, 12227-010 SP, Brazil \\
              Tel.: +55-12-32087214\\
              \email{edgard.evangelista@inpe.br}
           \and
           Jos\'{e} C N de Araujo \at
              Instituto Nacional de Pesquisas Espaciais -- Divis\~{a}o de Astrof\'{i}sica, Av. dos Astronautas 1758, S\~{a}o Jos\'{e} dos Campos, 12227-010 SP, Brazil \\
              Tel.: +55-12-32087223\\
              Fax: +55-12-32086811\\
              \email{jcarlos.dearaujo@inpe.br}
}

\date{Received: date / Accepted: date}

\maketitle

\begin{abstract}
Binary Systems are the most studied sources of gravitational waves. The mechanisms of emission and the behavior of the orbital parameters are well known and can be written in analytic form in several cases. Besides, the strongest indication of the existence of gravitational waves has arisen from the observation of binary systems. On the other hand, when the detection of gravitational radiation becomes a reality, one of the observed pattern of the signals will be probably of stochastic background nature, which are characterized by a superposition of signals emitted by many sources around the universe. Our aim here is to develop an alternative method of calculating such backgrounds emitted by cosmological compact binary systems during their periodic or quasiperiodic phases. We use an analogy with a problem of Statistical Mechanics in order to perform this sum as well as taking into account the temporal variation of the orbital parameters of the systems. Such a kind of background is of particular importance since it could well form an important foreground for the planned gravitational wave interferometers DECI-Hertz Interferometer Gravitational wave Observatory
(DECIGO), Big Bang Observer (BBO), Laser Interferometer Space Antenna (LISA) or Evolved LISA (eLISA), Advanced Laser
Interferometer Gravitational-Wave Observatory (ALIGO) and Einstein Telescope (ET).

\keywords{Gravitational waves \and Stochastic background \and Compact binaries}
\end{abstract}

\section{Introduction}
\label{intro}

Of all the theoretically possible sources of gravitational waves, the compact binary systems, that is, those systems whose components are black holes or neutron stars, are the most interesting, both from the theoretical and observational viewpoints. Concerning the theoretical aspects, it is possible to derive equations which hold even in the advanced stages of spiralling, while from the observational viewpoint, binary systems are among the most probable sources to have their gravitational radiation detected in the near future by the planned gravitational wave interferometers DECI-Hertz
Interferometer Gravitational wave Observatory (DECIGO), Big Bang Observer (BBO), Laser Interferometer Space
Antenna (LISA) or Evolved LISA (eLISA), Advanced Laser
Interferometer Gravitational-Wave Observatory (ALIGO)
(and VIRGO), and ET, as we will see later on.

On the other hand, concerning the study of the gravitational waves themselves, of special interest for us are the stochastic backgrounds, which are characterized by spectra spanning a wide range of frequencies, where we cannot distinguish individual sources, instead, we can distinguish only a ``smooth'' shape, characteristic for each type of physical process which created them.

Here we are interested in the background of gravitational waves generated by compact binary systems during their periodic or quasiperiodic phases in a cosmological scenario. The background considered here is similar to the confusion noise generated by Galactic binaries in the LISA or eLISA frequency band\cite{nelemans}.

Before proceeding, it is worth mentioning the well-known fact that other authors studied such a subject. For example, in a recent review paper, Regimbau\cite{regimbau11} considered the astrophysical gravitational wave stochastic background, where the background by the compact binaries is also considered. Recently, Zhu et al\cite{zhu11} considered the stochastic gravitational wave background from coalescing binary black holes and Rosado\cite{rosado} the gravitational wave background from compact binary systems in general. It is also worth mentioning some earlier works on this subject, including Farmer and Phinney\cite{farmer}, Regimbau and de Freitas Pacheco\cite{regimbau2} and Schneider et al\cite{ferrari}. Some ingredients considered by many of these authors include, for example, population synthesis and cosmic star formation histories.

Given that many authors studied the background generated by compact binaries, one could ask whether it is still worth studying such a subject. Since there is other possible approaches, such as the one considered in the present paper, there is still space for keeping studying such a subject.

Then, we shall present here an alternative method to calculate such a background, which is organized in steps. First, in Section \ref{section2}, we present some properties of the gravitational radiation emitted by binary systems such as frequencies, intensities and their relation with the orbital parameters; in Section \ref{section3}, we describe the population characteristics of the systems in the form of distribution functions involving orbital parameters such as orbital distances and masses of the components of the systems and the star formation rate. Once we have individual and population characteristics, we are ready to calculate the stochastic background, and in Sect. \ref{section4}, a method for this calculation is shown, where we adapted a method used in the solution of a simple problem of Statistical Mechanics in order to obtain the desired background spectrum; in Section \ref{section5}, we present and discuss some results obtained by using the method mentioned above, besides showing and justifying the numerical parameters we used in the calculation; in Section \ref{section6}, we consider the detectability of the spectra by the cross-correlation of pairs of LIGO and ET, and in Section \ref{conclusion}, we present some conclusions and perspectives. Finally, in order to give the reader more details, there is one Appendix at the end of the paper.

\section{Binary Systems and Gravitational Waves}
\label{section2}

In the present paper, we are concerned with systems in circular orbits, where the only harmonic present in the emitted radiation is the first one, whose frequency is twice the orbital frequency; for non null eccentricities, the higher harmonics will be in general important, in such a way that a fraction of the total power will be emitted at higher harmonics\cite{thorne,peters,kowa}.

For circular orbits, the expression for the total power emitted in gravitational waves by binary systems is given by\cite{weinberg}:

\begin{equation}
\frac{dE}{dt}=\frac{32G\Omega^{6}}{5c^{5}}\left(\frac{m_{1}m_{2}}{m_{1}+m_{2}}\right)^{2}r^{4}.
\label{pot}
\end{equation}
where $m_{1}$ and $m_{2}$ are the masses of the components of the system, $r$ is the orbital distance and $\Omega$ is the orbital angular frequency which can be written as a function of the orbital distance via Kepler's third law:
\begin{equation}
\label{kepler1}
\Omega^{2}=\frac{G(m_{1}+m_{2})}{r^{3}}.
\end{equation}

Besides, it is possible to put $\Omega$ as a function of time; with the aid of (\ref{pot}) and after some algebraic manipulation, we obtain
\begin{equation}
\frac{d\Omega}{dt}=\frac{96m_{1}m_{2}}{5c^{5}}G^{5/3}(m_{1}+m_{2})^{-1/3}\Omega^{11/3}
\label{frequ}
\end{equation}
whose integration yields the orbital frequency as a function of time, namely,
\begin{equation}
\Omega^{-8/3}=\Omega_{0}^{-8/3}-\frac{8}{3}K(t-t_{0}).
\label{freq}
\end{equation}

In (\ref{freq}), $\Omega_{0}$ is the initial frequency, $t_{0}$ is the initial time and $K$ is a constant given by
\begin{equation}
K=\frac{96m_{1}m_{2}}{5c^{5}}G^{5/3}(m_{1}+m_{2})^{-1/3}
\label{const}
\end{equation}

This expression, even though valid only for circular orbits, will be useful because the method developed in the present paper for the calculation of the stochastic background was first applied to circular orbits. Besides, it gives us a good physical insight on how the evolution of the systems takes place.

\section{Population Characteristics}
\label{section3}

Binary systems that contribute to the background may present orbital parameters such as semimajor axis and masses of the components that are quite different from system to system, which generally fill a continuous range of values. In general, it is possible to define mathematical functions which describe how the several values of a given parameter are distributed among the elements of the population. Such distribution functions are vital to the calculation of the background and may be found in the literature under several forms.

\subsection{Mass Distribution Function and the Star Formation Rate}

We are considering here the so called Salpeter distribution\cite{salpeter}, which is written as

\begin{equation}
\phi(m)=Am^{-(1+x)}
\label{salpeter1}
\end{equation}
where $x=1.7$ and $A=0.17$, with $\phi(m)$ obeying the following normalization condition
\begin{equation}
\int^{m_{f}}_{m_{i}}m\phi(m)dm=1.
\end{equation}
Here, $m_{f}$ and $m_{i}$ are the maximum and minimum values of the masses that stars can have, respectively.

We consider that stars can have minimum masses of $0.1\mbox{M}_{\odot}$ and maximum masses of $125\mbox{M}_{\odot}$\cite{ostlie}, where this interval is split in subdivisions: stars with masses between $0.1\mbox{M}_{\odot}$ and $8\mbox{M}_{\odot}$ give rise to white dwarfs at the final stage of their evolution; stars with masses in the interval between $8\mbox{M}_{\odot}$ and $25\mbox{M}_{\odot}$ originate neutron stars and stars more massive than $25\mbox{M}_{\odot}$ end their existences as black holes.

It is worth mentioning that Salpeter distribution is valid only for the masses of the progenitor stars, and when such progenitors become black holes or neutron stars, their masses change, so we must estimate the values of the masses of such remnants. For the mass of the neutron stars, we use a constant value of $1.4\mbox{M}_{\odot}$, according to the Chandrasekhar limit and the theories of formation of supernovae II and Ib$/$c\cite{ostlie}. For the remnant black hole mass distribution, we use (\ref{blackhole}), found in a recent study by \"{O}zel et al\cite{feryal}, namely

\begin{equation}
G_{\mbox{\scriptsize{bh}}}(m_{\mbox{\scriptsize{bh}}})=0.332\hspace{1pt}\mbox{exp}\left[-0.347(m_{\mbox{\scriptsize{bh}}}-7.8)^{2}\right],
\label{blackhole}
\end{equation}
where the masses are given in solar masses. Concerning the range of masses of black holes, we consider the range of $5\mbox{M}_{\odot}\leq m_{\mbox{\scriptsize{bh}}}\leq 20\mbox{M}_{\odot}$\cite{feryal,mandel}. However, we must emphasize that there are uncertainties in the determination of exact values of this interval of masses, so we chose values that seem suitable, bearing in mind the cited references. An interesting study concerning the black hole formation parameters can also be found in Postnov and Yungelson\cite{post}. Later on, we comment on how our results would be affected if a wider range of values were assumed.

However, the description of the star population is not complete without the star formation rate density (SFRD), and we use the function by Springel and Hernquist\cite{springel}, where the authors -- considering a $\Lambda$CDM cosmology in a structure formation scenario with the following cosmological parameters: $\Omega_{\mbox{\scriptsize{M}}}=0.3$, $\Omega_{\Lambda}=0.7$, Hubble constant $H_{0}=100h \,\mbox{km}\mbox{s}^{-1}\mbox{Mpc}^{-1}$ with $h=0.7$, $\Omega_{\mbox{\scriptsize{B}}}=0.04$ and a scale invariant power spectrum with index $n=1$, normalized to the abundance of rich galaxy clusters at present day ($\sigma_{8}=0.9$) -- found a star formation rate covering a range of redshifts from $z=0$ to the ``dark ages'' ($z\sim 20$). Such a function is given by
\begin{equation}
\dot{\rho}_{\ast}(z)=\rho_{m}\frac{\beta e^{\alpha(z-z_{m})}}{\beta-\alpha+\alpha e^{\beta(z-z_{m})}}
\label{starfor}
\end{equation}
where $\alpha=3/5$, $\beta=14/15$, $z_{m}=5.4$, and with $\rho_{m}=0.15\mbox{M}_{\odot}\mbox{yr}^{-1}\mbox{Mpc}^{-3}$ fixing the normalization. Further, it is worth mentioning that the authors used hydrodynamic simulations, taking into account radiative heating and cooling of gas, supernova feedbacks and galactic winds.

Even though (\ref{starfor}) gives information about the star formation up to $z=20$, we must point out that the observations can give information from at most $z\sim 5$. Therefore, uncertainties are inherent in any modeling above a given redshift.  Also, the star formation rate adopted in this work refers to the first population II stars formed; therefore it is consistent with the Salpeter IMF adopted.

On the other hand, in order to see how the spectra depend on this important ingredient, we also
performed calculations using other SFRDs. In particular, we considered four
additional star formation rates\cite{zhufan}. Three of
them are given by
\begin{equation}
\dot{\rho}_{\mbox{\textasteriskcentered}}(z)_{i}=1.67C_{i}h_{65}F(z)G_{i}(z)\mbox{M}_{\odot}\mbox{yr}^{-1}\mbox{Mpc}^{-3}
\label{sfr1}
\end{equation}
with $i=1,2,3$ denoting each case, $h_{65}=h/0.65$, $h=0.7$, $F(z)=[\Omega_{M}(1+z)^{3}+\Omega_{\Lambda}]^{1/2}/(1+z)^{3/2}$, $C_{1}=0.3$, $C_{2}=0.15$, $C_{3}=0.2$, $\Omega_{\Lambda}=0.7$, $\Omega_{M}=0.3$ and the functions $G_{i}(z)$ given by
\begin{eqnarray}
G_{1}(z)&=&\frac{\mbox{e}^{3.4z}}{\mbox{e}^{3.8z}+45} \\
G_{2}(z)&=&\frac{\mbox{e}^{3.4z}}{\mbox{e}^{3.4z}+22} \\
G_{3}(z)&=&\frac{\mbox{e}^{3.05z-0.4}}{\mbox{e}^{2.93z}+15}
\end{eqnarray}
The fourth one has the following form:

\begin{equation}
\dot{\rho}_{\mbox{\textasteriskcentered}}(z)=h\frac{0.017+0.13z}{1+(z/3.3)^{5.3}}\mbox{M}_{\odot}\mbox{yr}^{-1}\mbox{Mpc}^{-3}
\label{sfr2}
\end{equation}

It is worth mentioning that we will not consider here a complete description of these additional SFRD's. We refer the reader to the paper by  Zhu et al\cite{zhufan} for a detailed discussion concerning this issue.

\begin{figure}
\rotatebox{-90}{\includegraphics[width=0.30\textwidth]{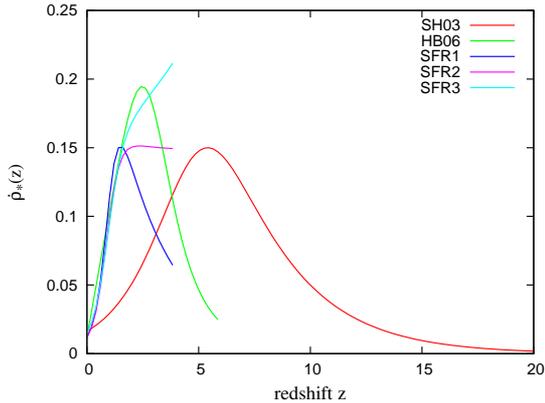}}
\caption{The SFRD's adopted in the present paper. See the text for details}
\label{sfrds}
\end{figure}

Notice that the three first functions describe formation rates for redshifts of up to $z\approx 4$ and the fourth one is valid for redshifts of up to $z\approx 6$. Besides, we use the following notation\cite{zhufan}: for the rates given by (\ref{sfr1}) we use the labels SFR$1$, SFR$2$ and SFR$3$ and for (\ref{sfr2}) we use HB$06$; the formation rate by Springel and Hernquist is written as SH$03$. Figure \ref{sfrds} shows the SFRDs adopted in the present paper.

\subsection{The Binary Formation Rates}

Before proceeding, it is necessary to know the rate of formation of compact binary objects; that is, given a population of progenitor stars, we need to know the fraction of stars which generate double compact binaries. As we may suppose, such fractions may assume different values because, from the formation of a population of stars to the final configuration (which is characterized by a population containing main sequence stars, neutron stars, black holes and binaries formed from these objects), several processes may affect the final number of compact binaries present. For example, if a component of a binary system gives rise to a supernova, the explosion may break the system. Further, note that not all progenitor stars form binary systems, and not all stars will generate black holes or neutron stars.

For the formation rate of double neutron star (DNS) systems, we consider the mass fraction $\lambda_{\mbox{\scriptsize{dns}}}$ that is converted into DNS systems\cite{regimbau2}, which is given by

\begin{equation}\label{}
    \lambda_{\mbox{\scriptsize{dns}}}=\beta_{\mbox{\scriptsize{ns}}}f_{p}\Phi_{\mbox{\scriptsize{ns}}}
\end{equation}
where $\beta_{\mbox{\scriptsize{ns}}}$ is the fraction of binary systems that survive to the second supernova event, $f_{p}$ gives us the fraction of massive binaries (that is, binary systems where both components could generate a supernova event) formed inside the whole population of stars, and $\Phi_{\mbox{\scriptsize{ns}}}$ is the mass fraction of neutron star progenitors that, in our case and using (\ref{salpeter1}), is given by
\begin{equation}\label{}
    \Phi_{\mbox{\scriptsize{ns}}}=\int^{25}_{8}\phi(m)dm.
\end{equation}
Numerically, we have $\beta_{\mbox{\scriptsize{ns}}}=0.024$, $f_{p}=0.136$ and $\Phi_{\mbox{\scriptsize{ns}}}=5.97\times 10^{-3}\mbox{M}_{\odot}^{-1}$.

Finally, the binary formation rate for DNS systems is given by
\begin{equation}
n_{\mbox{\scriptsize{dns}}}(z)=\lambda_{\mbox{\scriptsize{dns}}}\dot{\rho}_{\ast}(z).
\label{bin_form}
\end{equation}

For black hole-neutron star (BHNS) and binary of black holes (BBH) systems, we consider that the population of compact binaries is composed by DNS systems ($61\%$), BBH systems ($30\%$) and BHNS systems ($9\%$)\cite{belczynski}. Therefore, we may use these proportions to estimate the mass fraction for BHNS and BBH systems.

\subsection{The Semimajor Axis Distributions}

\begin{figure}
\rotatebox{-90}{\includegraphics[width=0.30\textwidth]{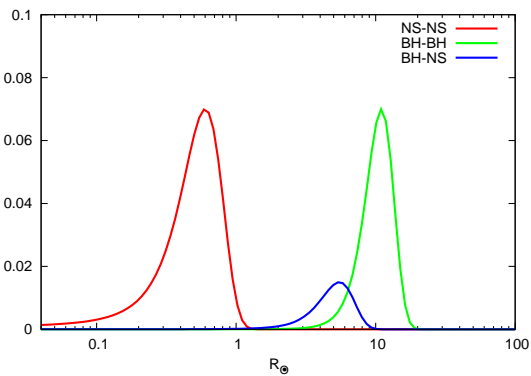}}
\caption{Distribution functions of orbital separations of compact binaries}
\label{gaussian}
\end{figure}

For the distribution function of the semimajor axis, we followed Belczynski et al\cite{belczynski}, who made a very comprehensive study of the compact binary systems. In particular, we used the results present in Fig. $4$ of that paper. To deal more easily with the distribution functions obtained by Belczynski et al, we fit them with gaussian functions. A close comparison between our functions (see Fig.~\ref{gaussian}) and Fig. $4$ of the cited paper shows that the fits work very well.

The semimajor axis distributions are then written as follows:
\begin{equation}
f(r)=C\hspace{1pt}\mbox{exp}\left[\frac{-(r-\bar{r})^{2}}{2\sigma^{2}}\right]
\label{distr1}
\end{equation}
where $r$ is the semimajor axis, which is given in solar radius units, and the parameters $C$, $\bar{r}$ and $\sigma$
are given in Table \ref{gauss}.

\begin{table}
\caption{Parameters of the distribution functions of the orbital distances}
\label{gauss}
\begin{tabular}{llll}
\hline\noalign{\smallskip}
system & $C$ & $\bar{r}(\mbox{R}_{\odot})$ & $\sigma(\mbox{R}_{\odot})$ \\
\noalign{\smallskip}\hline\noalign{\smallskip}
DNS &$0.070$ &$\hphantom{0}0.6$ &$0.2$\\
BHNS &$0.015$ &$\hphantom{0}5.5$ &$1.5$\\
BBH &$0.070$ &$11.0$ &$2.5$\\
\noalign{\smallskip}\hline
\end{tabular}
\end{table}

It is worth noting that the above distribution functions obey the following normalization condition\cite{belczynski}:

\begin{equation}
\int^{\infty}_{-\infty}[f_{\mbox{\scriptsize{dns}}}(r)+f_{\mbox{\scriptsize{bhns}}}(r)+f_{\mbox{\scriptsize{bbh}}}(r)]dr=1
\end{equation}

In addition, it is useful to have information about the orbital periods or frequencies of the systems. A convenient way to achieve this is to plot the distribution of the initial frequencies. As separations and frequencies are closely related by Kepler's third law, it is straightforward to obtain the distribution of the frequencies from (\ref{distr1}) (see the next Subsection).

\subsection{Time Evolution of $f(r)$}

A remarkable feature of binary systems is the time evolution of the orbital distance due to the emission of gravitational radiation. For the systems considered in this paper, such variations are very important, especially at the stages near the coalescence, so the distribution functions shown above must be modified in order to take into account the time as an independent variable. First, let us consider the distribution given by (\ref{distr1}), which will be modified with the aid of (\ref{freq}), valid for circular orbits.

Before including the time variable in (\ref{distr1}), it is convenient to put this function in terms of the frequency, by means of (\ref{kepler1}). So, changing the variables via

\begin{equation}
f(r)dr=g(\Omega)d\Omega
\label{transf}
\end{equation}
we get after some algebra

\begin{eqnarray}\nonumber
g(\Omega)&=&\frac{2C}{3}\\
&&\times\left[G(m_{1}+m_{2})\right]^{1/3}\Omega^{-5/3}\mbox{exp}\left[\frac{-(r-\bar{r})^{2}}{2\sigma^{2}}\right]
\label{distr2}
\end{eqnarray}

\begin{figure}
\rotatebox{-90}{\includegraphics[width=0.30\textwidth]{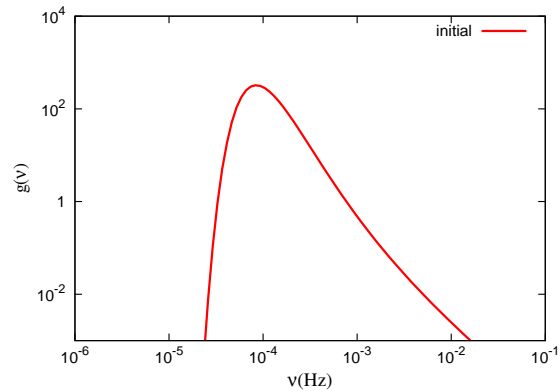}}
\caption{Distribution $g(\nu)$ for NSNS systems}
\label{evolfreq}
\end{figure}

Now we use (\ref{freq}) and perform a change of variables in (\ref{distr2}), writing
\begin{equation}
g(\Omega_{0})d\Omega_{0}=H(\Omega)d\Omega
\label{distr3}
\end{equation}
where $\Omega_{0}$ is the initial frequency, which was associated with the variable $\Omega$ in (\ref{distr2}). So, performing the derivative $\frac{d\Omega_{0}}{d\Omega}$ of (\ref{freq}) we have
\begin{equation}
\frac{d\Omega_{0}}{d\Omega}=\left(\frac{\Omega_{0}}{\Omega}\right)^{11/3}
\end{equation}
which, substituting in (\ref{distr3}), gives after some algebra

\begin{eqnarray}\nonumber
H(\Omega)&=&\frac{2C}{3}\left[G(m_{1}+m_{2})\right]^{1/3}\\
&&\times\Omega^{-11/3}\Omega^{2}_{0}\mbox{exp}\left[\frac{-(r-\bar{r})^{2}}{2\sigma^{2}}\right]
\label{dist1}
\end{eqnarray}
where
\begin{equation}
\Omega_{0}=\left[\Omega^{-8/3}+\frac{8}{3}K(t-t_{0})\right]^{-3/8}
\end{equation}
and $K$ is given by (\ref{const}). We should do a further coordinate transformation, writing $H(\Omega)$ as a function of the emitted frequency $\nu$ instead of the angular frequency $\Omega$. Such a transformation, given by $\nu=\Omega/\pi$, is trivial, and all the equations used from now on will be written as a function of $\nu$. As an example, Fig.~\ref{evolfreq} shows the initial distribution $g(\nu)$ for DNS systems.

Note that $r$ and $\bar{r}$ were maintained in the above equations just for simplifying their presentations; they are effectively substituted by the appropriate set of variable transformations.

It is worth noting that Hils et al\cite{hils} followed a similar procedure; namely, they started with a given initial distribution function and evolved it as a function of time.

\subsection{Cosmological Parameters}

The cosmological model adopted here, which is the $\Lambda$CDM, has the volume element given by
\begin{equation}
\frac{dV}{dz}=4\pi \left(\frac{c}{H_{0}}\right)r_{z}^{2}F(\Omega_{\mbox{\scriptsize{M}}},\Omega_{\mbox{\scriptsize{$\Lambda$}}},z),
\end{equation}
where
\begin{equation}
F(\Omega_{\mbox{\scriptsize{M}}},\Omega_{\Lambda},z)=\frac{1}{\sqrt{(1+z)^{2}(1+\Omega_{\mbox{\scriptsize{M}}}z)-z(2+z)\Omega_{\Lambda}}},
\end{equation}
and the comoving distance given by
\begin{equation}
r_{z}=\frac{c}{H_{0}\sqrt{\vert\Omega_{\mbox{\scriptsize{k}}}\vert}}S\left[\sqrt{\vert\Omega_{\mbox{\scriptsize{k}}}\vert}
\int_{0}^{z}F(\Omega_{\mbox{\scriptsize{M}}},\Omega_{\Lambda},z')dz'\right],
\end{equation}
where the density parameters (which should not be confused with the frequency $\Omega$ in the previous sections) obey the following relations:
\begin{equation}
\Omega_{\mbox{\scriptsize{M}}}=\Omega_{\mbox{\scriptsize{DM}}}+\Omega_{\mbox{\scriptsize{B}}}\:\:\:\mbox{and}\:\:\:\Omega_{\mbox{\scriptsize{M}}}+\Omega_{\Lambda}+\Omega_{\mbox{\scriptsize{k}}}=1
\end{equation}
The subscripts M, DM, B, $\Lambda$ and k refer to matter, dark matter, baryonic matter, cosmological constant and curvature, respectively. Furthermore, the function $S$ is defined by
\begin{displaymath}
S(x)=\left\{\begin{array}{rc}
\mbox{sin}(x) &\hspace{3pt}\mbox{closed universe}\\
x     &\hspace{3pt}\mbox{flat universe}\\
\mbox{sinh}(x)&\hspace{3pt}\mbox{open universe}
\end{array}\right.
\end{displaymath}

In the present paper we consider the case of a flat Universe (i.e. $\Omega_{\mbox{\scriptsize{k}}}=0$), where the cosmological parameters
were presented in the previous section.

\section{The Stochastic Background}
\label{section4}

The stochastic background of gravitational waves generated by compact binaries will be calculated by means of\cite{araujo05}
\begin{equation}
h_{\mbox{\scriptsize{BG}}}^{2}=\frac{1}{\nu_{\mbox{\scriptsize{obs}}}}\int h_{\mbox{\scriptsize{source}}}^{2}dR
\label{bg}
\end{equation}
where $h_{\mbox{\scriptsize{BG}}}$ represents the dimensionless amplitude of the spectrum, $h_{\mbox{\scriptsize{source}}}$ is the amplitude of the emitted radiation of a single source,  $\nu_{\mbox{\scriptsize{obs}}}$ is the observed frequency, and $dR$ is the differential rate of generation of gravitational radiation. The differential rate of production of gravitational waves is derived by using a statistical approach. One of the aims of such a derivation was to take in to account the change in the gravitational wave frequency emitted by binary systems that are particularly important for the system emitting at the high band of the spectra.

In the present case $h_{\mbox{\scriptsize{source}}}$ is given by\cite{evans,thorne}

\begin{eqnarray}\nonumber
h_{\mbox{\scriptsize{source}}}&=&7.6\times 10^{-23}\\
&&\times\left(\frac{\mu}{M_{\odot}}\right)\left(\frac{M}{M_{\odot}}\right)^{2/3}\left(\frac{1\mbox{Mpc}}{d_{\mbox{\scriptsize{L}}}}\right)\left(\frac{\nu}{1\mbox{Hz}}\right)^{2/3}
\label{source}
\end{eqnarray}
where $\mu$ is the reduced mass of the system, $M$ is the total mass, and $d_{\mbox{\scriptsize{L}}}$ is the luminosity distance. Besides, the differential rate $dR$ in (\ref{bg}) for the case of BBH systems may be written as
\begin{equation}
    dR_{\mbox{\scriptsize{bbh}}}=\frac{dR}{dz}G_{\mbox{\scriptsize{bh}}}(m_{1})G_{\mbox{\scriptsize{bh}}}(m_{2})dzdm_{1}dm_{2}
\label{taxa}
\end{equation}

Notice that we consider in the above equation the most general case, where the masses of the components of the systems are not constant, which in our case refers to the systems which are formed by two black holes obeying the distribution function given by (\ref{blackhole}). For the case where one or both components are neutron stars, one or both distributions for the masses in (\ref{taxa}) will be considered as a Dirac delta, since we are considering that all neutron stars have the same  masses. Further, note that, in order to simplify the notation, we have written the masses of the black holes as $m_{1}$ and $m_{2}$. In addition, the term $dR/dz$ will be determined in a subsequent section.

Still concerning (\ref{bg}), it is worth mentioning that the starting point of its derivation comes from an energy flux equation. This equation was first derived in a paper by de Araujo et al\cite{araujo00} and was used in their various articles. In particular, in a subsequent paper\cite{araujo05} they gave a more detailed derivation of this equation, showing its robustness. Also, although apparently simple, it contains the correct and necessary ingredients to calculate the background of gravitational waves for a given type of source.

It is also important to call attention to a particular issue considered in the cited paper\cite{araujo05}. They raised a discussion regarding the inclusion or not of an additional $(1+z)$ factor in the denominator of the integrand of (\ref{bg}), which many authors consider to take into account the time dilation due to the expansion of universe. These authors argued that when one uses the luminosity distance in the calculation of the gravitational wave amplitude $h_{\mbox{\scriptsize{source}}}$, this distance already contains all necessary $(1+z)$ factors. This is in fact a controversial issue. In practice, however, the inclusion of the additional time dilation factor does not modify significantly the amplitude of the gravitational wave background. Therefore, in the present paper we included the $(1+z)$ factor in the
denominator of the integrand of (\ref{bg}).

\subsection{The Statistical Problem}

Basically, the problem of determining the stochastic background comes down to the counting of the number of binary systems that, at a given instant of time, emit gravitational waves at a given frequency. Specifically, when one is dealing with high frequencies (about $1$Hz or more), the systems evolve very fast, and the frequencies of the emitted radiation vary in a small timescale, making the calculation of the systems which have the same frequency at a given moment of time more difficult. However, the problem can be handled by adopting a particular way of modelling the behavior of the population of binary systems; let us idealize a one-dimensional ``space'' where the values of the coordinates are the values of the emitted frequencies.

In this context, the evolving binary systems can be represented as points moving in the direction of increasing values of the coordinate (emitted frequency). At this point, we can note that those points have a behavior similar to a gas inside a chamber, where the particles have a certain spatial distribution and move towards a given wall of the chamber at different velocities. So, using this analogy, we can adapt some techniques of statistical mechanics in order to develop a method to calculate the background.

\begin{figure}
\rotatebox{0}{\includegraphics[width=0.30\textwidth]{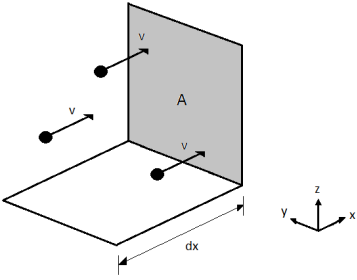}}
\caption{The volume element $dV$ adjacent to $A$}
\label{excent}
\end{figure}

Let us consider a gas in thermal equilibrium contained in a rectangular chamber of cross section (relatively to the $x$ axis) equal to $A$. An elementary question concerning such a configuration is to calculate how many particles (or molecules) reach a given surface (perpendicular to the $x$ axis) of area $A$ in a time interval $dt$; that is, the aim is to calculate the flux $F$ of particles through a given area. Discussions on this question may be found in the literature, and we follow a particular way of handling the problem\cite{jackson}; let then $Adx=dV$ be a differential volume adjacent to the surface $A$, according to Fig.~\ref{excent}.

If we consider that the particles are distributed in a homogeneous way inside the chamber, the fraction of particles inside the volume $dV$ is given by
\begin{equation}
    dn=\frac{dV}{V}=\frac{Adx}{V}
\label{german}
\end{equation}
where $V$ is the total volume of the chamber, but using the velocity of a given particle at the $x$ direction we have:
\begin{equation}
    dV=Av_{x}dt
\label{volume}
\end{equation}

In this way, all the particles inside the volume $dV$ will reach the surface $A$ in a time interval $dt$. In order to calculate the total number of particles that hit the area $A$, we should integrate over the velocities $v_{x}$, using a given velocity distribution function, which we write as $\eta(v)$. So, considering that $\eta(v)dv^{3}$ is the total number of particles at a given interval $dv^{3}$, we get
\begin{equation}
	  d\mu=\frac{Av_{x}dt}{V}\eta(v)d^{3}v
\label{fracao}
\end{equation}
where $d\mu$ is the number of particles in $dV$, which have velocities in the interval $d^{3}v$.
Now, we should integrate only over the positive values of $v_{x}$, because we are considering only
particles that are moving towards the surface $A$, so we get the number of particles per area $A$ and time interval $dt$ (the desired flux $F$):
\begin{equation}
F=\frac{1}{V}\int^{\infty}_{0}dv_{x}\int^{\infty}_{-\infty}dv_{y}\int^{\infty}_{-\infty}dv_{z}v_{x}\eta(v)
\end{equation}

This is sufficient for the solution of the statistical problem, but we need to make two modifications in the above derivation in order to adapt it to the calculation of the stochastic background. First, note that the cross section of the chamber is constant, so (\ref{german}) may be rewritten as

\begin{equation}
    dn=\frac{Adx}{AL}=\frac{dx}{L}
\end{equation}

\noindent where $L$ is the total length of the chamber. Thus (\ref{fracao}) is rewritten as

\begin{equation}
    d\mu=\frac{dx}{L}\eta(v)d^{3}v.
\end{equation}

A second modification is to consider a more general case, where the spatial distribution is not constant. So, it is reasonable to write the expression above as
\begin{equation}
    d\mu=\psi(x)\eta(v)d^{3}vdx
\label{back1}
\end{equation}
where $\psi(x)dx$ is the fraction of particles at the interval $dx$. If the spatial distribution obeys a function which depends only on $x$, which we call $\varphi(x)$, we can write the function $\psi(x)$ in the following way
\begin{equation}
    \psi(x)=\frac{\varphi(x)}{\int \varphi(x)dx}
\label{back2}
\end{equation}
for the case where $\varphi(x)$ is not normalized. With these two considerations at hand, the problem is ready to be used in the calculation of $dR$.

\subsection{Calculation of $dR$}

Before proceeding, it is worth stressing that the derivation of $dR$ presented here refers to the source frame.
Therefore, $\nu$ in the equations below is the emitted frequency, and the time derivatives are also taken with respect to
the source frame.

The calculation of $dR$ by means of the statistical approach shown above is done by performing two changes: instead of using the spatial distribution $\varphi(x)$, let us use a frequency distribution $\varphi(\nu)$ and change the velocity distribution for a distribution in a new variable $\upsilon_{\nu}$, which we shall define by
\begin{equation}
    \upsilon_{\nu} =\frac{d\nu}{dt}
\end{equation}
that is, the time derivative of the frequency (in analogy to the definition of velocity, which establishes
the time variation of the spatial coordinate). Once this is done, we need to find out the forms of these
two new functions.

So, the function $\varphi(\nu)$ may be defined as follows:
\begin{equation}
\label{inte}
   \varphi(\nu) =\int n_{\mbox{\scriptsize{bin}}}(t_{0})H(t,t_{0},\nu)dt_{0}
\end{equation}
where $t_{0}$ represents the instant of the birth of the systems, $ n_{\mbox{\scriptsize{bin}}}(t_{0})$ is the formation rate of binary systems (where the subindex $bin$ is to be substituted by dns, bhns and bbh in each case) and $H(t,t_{0},\nu)$ is given by (\ref{dist1}) and represents the fraction of systems that begins to exist at the instant $t_{0}$ and that have frequencies in a given interval $d\nu$. Note that here, for sake of notation, we explicit the dependence of (\ref{dist1}) on $t$ and $t_{0}$. Besides, we refer the reader to Appendix \ref{appenA} to see the derivation of (\ref{inte}). Further, note that $z_{0}$ and $t_{0}$ are related to each other by $t_{0}=2H_{0}^{-1}[1+(1+z_{0})^{2}]^{-1}$\cite{timeredshift}.

The distribution $\eta(\upsilon_{\nu})$ will have a peculiar form. First, we need to note that from (\ref{frequ}) we have
\begin{equation}
\upsilon_{\nu}\equiv \frac{d\nu}{dt}\propto \nu^{\frac{11}{3}}
\end{equation}
Then, we conclude that there will be just one value of $\upsilon_{\nu}$ for each value of $\nu$, which allows us to write $\eta(\upsilon_{\nu})$ as a Dirac delta function \\
\begin{equation}
\label{xi}
    \eta(\upsilon_{\nu})=N\delta(\upsilon_{\nu}-\upsilon_{\nu,p})
\end{equation}

where $N$ is the total number of systems and $\upsilon_{\nu,p}$ is the particular value of $\upsilon_{\nu}$ corresponding to each frequency $\nu$.

Now we can rewrite (\ref{back1}) as \\
\begin{equation}
    d\mu=\left(\frac{\varphi(\nu)d\nu}{\int \varphi(\nu)d\nu}\right)\eta(\upsilon_{\nu})d\upsilon_{\nu}
\end{equation}
where we used (\ref{back2}) with the variable $x$ substituted by $\nu$. Now, noting that the denominator of the term between parenthesis is the total number of systems, using the function given by (\ref{xi}) and changing the differential $d\nu$ by means of the chain rule, we get
\begin{equation}
d\mu=\left(\frac{\varphi(\nu)\frac{d\nu}{dt}dt}{N}\right)N\delta(\upsilon_{\nu}-\upsilon_{\nu,p})d\upsilon_{\nu}
\end{equation}
By integrating over $\upsilon_{\nu}$ and rearranging the result, we obtain
\begin{equation}
R=\varphi(\nu)\frac{d\nu}{dt}
\end{equation}
where $R$ is number or systems by time interval $dt$. Recalling that this rate $R$ is per comoving volume, we may write explicitly
\begin{equation}
\label{bg4}
\varphi(\nu)\frac{d\nu}{dt}\equiv \frac{dR}{dV}
\end{equation}
and (\ref{bg}) assumes the form
\begin{equation}
\label{bg2}
h_{\mbox{\scriptsize{BG}}}^{2}=\frac{1}{\nu_{\mbox{\scriptsize{obs}}}}\int h_{\mbox{\scriptsize{source}}}^{2}\frac{dR}{dV} \frac{dV}{1+z}.
\end{equation}

Further, using this amplitude we can derive a largely used quantity, the so called spectral amplitude, which is given by:
\begin{equation}
S_{h}=\frac{h_{\mbox{\scriptsize{BG}}}^{2}}{\nu_{\mbox{\scriptsize{obs}}}}.
\end{equation}

It is worth noting that we use the classical (i.e., nonrelativistic) approach in the above statistical problem. There are two reasons for this choice: first, we should note that even in the late stages of evolution, the components of the binary systems have low orbital velocities (i.e., nonrelativistic velocities), and second, in all stages of evolution, the time variation of the orbital frequency ($\upsilon_{\nu}$) is not fast enough, and therefore, it is not necessary to consider special relativistic corrections when one makes the analogy between $\upsilon_{\nu}$ and the velocity of the particles. Concerning general relativistic effects, the well-known quadrupole formula is more than enough. This formula is (implicitly) used in the calculation of the time evolution of the orbital frequency (see, e.g., \cite{rosado}).

Besides, we refer the reader to the paper~\cite{efde}, where we show the (first) simple example of application of the method described here; namely, we calculate the stochastic background generated by double neutron star systems, considering an uniform orbital period distribution.

\section{Results and Discussions}
\label{section5}

One of the aims of the present paper is to investigate whether the background of gravitational waves by the cosmological compact binary systems could in principle generate a ``confusion noise" for the spatial and terrestrial interferometers; in the same way the Galactic binary systems do for LISA\cite{nelemans}. If such confusion noise really exists, it must necessarily be taken into account since it could well dominate the sensitivity curves of the detectors.

\subsection{Setting the Maximum and Minimum Frequencies}

For DNS systems, we consider for the maximum frequency of the gravitational waves the emitted value of $\approx 900$Hz\cite{isco}; for BHNS systems, we use (\ref{kepler}) in order to determine the maximum frequency
\begin{equation}
\nu_{\mbox{\scriptsize{max}}}=\left[\frac{G(m_{\mbox{\scriptsize{ns}}}+m_{\mbox{\scriptsize{bh}}})}{\pi^{2}r_{\mbox{\scriptsize{isco}}}^{3}}\right]^{1/2}
\label{kepler}
\end{equation}
where $m_{\mbox{\scriptsize{ns}}}=1.4\mbox{M}_{\odot}$, $m_{\mbox{\scriptsize{bh}}}=5.0\mbox{M}_{\odot}$ and $r_{\mbox{\scriptsize{isco}}}$ is the innermost stable circular orbit (ISCO) of the black hole\cite{kenyon}, given by
\begin{equation}
r_{\mbox{\scriptsize{isco}}}=\frac{6Gm_{\mbox{\scriptsize{bh}}}}{c^{2}}
\end{equation}

For BBH systems, the maximum frequency is given by\cite{marassi}

\begin{equation}
\nu_{\mbox{\scriptsize{max}}}=\frac{c^{3}}{G}\frac{a_{o}\eta^{2}+b_{0}\eta+c_{0}}{\pi M}
\label{merg}
\end{equation}
where $\eta=m_{\mbox{\scriptsize{bh}}_{1}}m_{\mbox{\scriptsize{bh}}_{2}}/M^{2}$ is the symmetric mass ratio, $M$ is the total mass and the polynomial coefficients are $a_{0}=2.974\times 10^{-1}$, $b_{0}=4.481\times 10^{-2}$ and $c_{0}=9.556\times 10^{-2}$.

The values of the minimum frequencies do not play a decisive role in the calculation; that is, they do not affect the shapes of the spectra. Therefore, we considered the value of $10^{-6}-10^{-5}$Hz for the three families of compact systems. This choice is suggested
by analyzing the behavior of (\ref{distr2}) (see also Fig.~\ref{evolfreq}). As can be seen,
the number of systems emitting below $10^{-6}-10^{-5}$Hz is negligle as compared with those emitting around $\sim 10^{-4}-10^{-3}$Hz.

\subsection{The spectra}

\begin{figure}
\rotatebox{-90}{\includegraphics[width=0.30\textwidth]{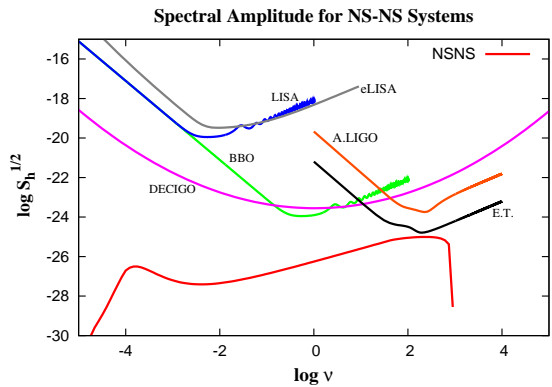}}
\caption{Spectral amplitude for DNS systems and the sensitivity curves of some interferometric detectors}
\label{espectro1}
\end{figure}

\begin{figure}
\rotatebox{-90}{\includegraphics[width=0.30\textwidth]{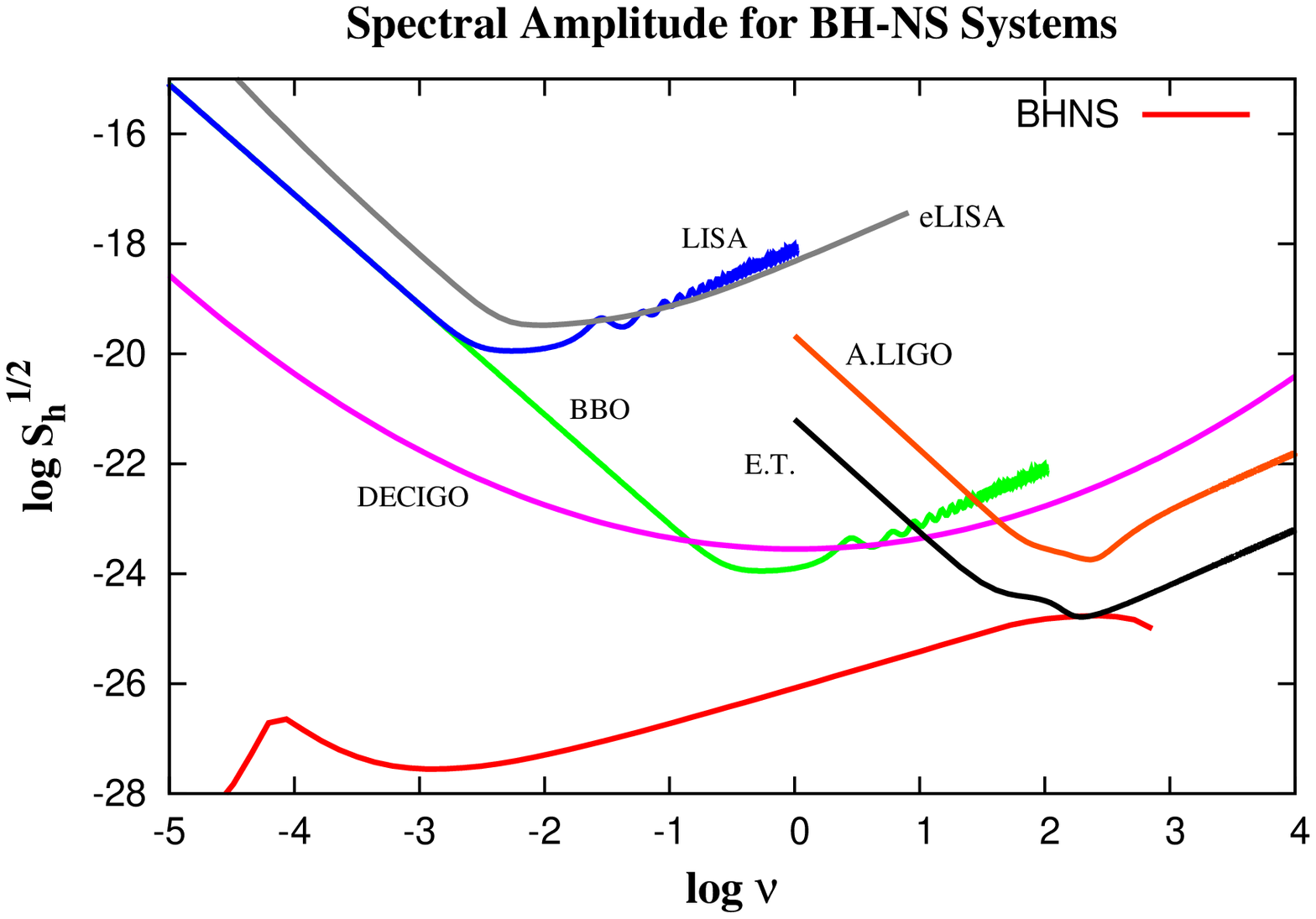}}
\caption{Spectral amplitude for BHNS systems and the sensitivity curves of some interferometric detectors}
\label{espectro2}
\end{figure}

\begin{figure}
\rotatebox{-90}{\includegraphics[width=0.30\textwidth]{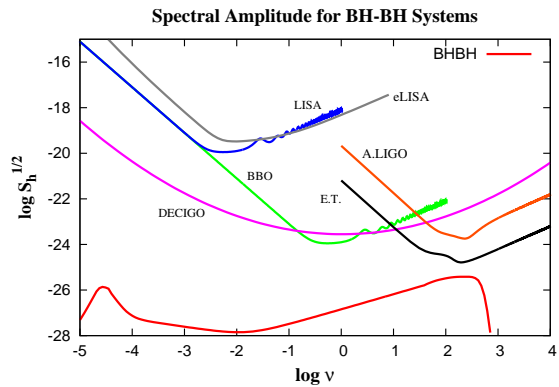}}
\caption{Spectral amplitude for BBH systems and the sensitivity curves of some interferometric detectors}
\label{espectro3}
\end{figure}

Figures \ref{espectro1}, \ref{espectro2} and \ref {espectro3} show the spectra for DNS, BHNS and BBH systems, respectively, which are compared to the sensitivity curves of LISA, eLISA\cite{elisa}, BBO\cite{bbo}, DECIGO\cite{decigo}, ET\cite{satya} and ALIGO\cite{satya}. The sensitivity curve for LISA may be found at \\ \url{www.srl.caltech.edu/~shane/sensitivity/}.

When observing the curves for the DNS and BBH systems, we note that LISA, eLISA, BBO, DECIGO, ALIGO and ET would not be affected by such a background. BHNS binaries would be marginally important for ET at the frequencies of highest sensitivity.

When observing the shape of the spectra, we note that there is a cutoff frequency, which in the present case is equal to the maximum frequency of the gravitational waves allowed for each type of system. Beyond this maximum frequency, the systems reach the chirp phase that is characterized by a particular pattern of emission, but we are not considering such a phase here. In fact, the chirp and coalescing phases will be considered in forthcoming papers.

It is also interesting to see the influence of the initial distribution function of frequencies on the spectra. As the distributions for orbital separations and frequencies are closely related, we can achieve this objective by handling the distribution of the initial separations. So, as an example, we calculated the spectra for DNS systems, considering the hypothetical cases where for the standard deviation $\sigma$ of the initial distribution given by (\ref{distr1}) gives the values $2\sigma$ and $\sigma/2$. As expected, the semimajor axis distribution has little influence on the spectra for frequencies much shorter than the smallest typical birth period.

Our calculations also show that the resulting spectra do not depend significantly on the SFRDs adopted. There are two main reasons for this result. The first one has to do with the fact that the sources located at $z>5$ do not significantly affect the resulting spectra. The second reason is related to the fact that the SFRDs for $z<5$ are similar.

Besides, through (\ref{source}), we observe the influence of the masses of the systems. For comparison, let us consider a  BBH system where both components have the maximum value of mass, but considering two cases: in the first case considering that the maximum mass is of $\sim 60\mbox{M}_{\odot}$\cite{woosley}, and in the second case, using the more realistic value of $20\mbox{M}_{\odot}$ as an upper limit (remembering that this is the value considered in the present paper). The amplitude given by (\ref{source}) in the first case is $\sim 6$ greater than the amplitude given by the second case. As a result, the amplitude of the background would also be higher.

Notice that the shapes of the spectra are different. The reason for this is the following. For a given semimajor axis, DNS systems emit gravitational waves at the same frequency, since we consider that all neutron stars have the same masses. For BBH (BHNS) systems, however, the emitted frequencies depend on the black holes masses that, contrary to the neutron stars, assume different values since they follow a distribution function.

It is worth paying attention to some similar studies found in the literature, since they could well consider different approaches. For example, in Schneider at al\cite{ferrari} the spectra were calculated for binary systems formed by neutron stars, black holes and white dwarfs at the early stages of spiralling. These authors used a different method, based on binary population synthesis programs. The differences in the results have to do with the fact that in their calculations, the number of systems that contributes to the lowest frequencies is much greater than that of those that contribute to the highest frequencies. Furthermore, these authors considered lower values for the minimum frequencies than those we are using.

In Regimbau\cite{regimbau}, the stochastic background generated by DNS systems in the frequency band of LISA is considered, where the author used a particular probability distribution\cite{dns} in order to generate a population of DNS systems. Here, the author considered the time evolution of the orbital parameters in a different way we do; actually, she considered the difference between the redshift of the birth of the systems $z_{\mbox{\scriptsize{b}}}$ and the redshift of the emission $z_{\mbox{\scriptsize{e}}}$ of gravitational waves, where $z_{\mbox{\scriptsize{b}}}$ and $z_{\mbox{\scriptsize{e}}}$ are related to each other by the frequency of the emitted waves.

\section{Cross Correlation of Detectors}
\label{section6}

The predicted spectra of the present paper cannot be detected by single interferometric detectors such as the ones cited above, but a putative detection of these spectra can occur by means of a suitable correlation of two or more detectors\cite{michelson,romano,allen}. More specifically, we will consider the cross correlation for pairs of Initial LIGOs (ILIGO), Enhanced LIGOs (ELIGO), ALIGOs and ETs.

For a cross correlation of two interferometric detectors, we can quantify the detectability of a stochastic background by calculating the signal-to-noise ratio (S$/$N), which in this case is given by\cite{allen}:

\begin{equation}
(S/N)^{2}=\left[\left(\frac{9H_{0}^{4}}{50\pi^{4}}\right)T\int \frac{\gamma^{2}(\nu)\Omega_{\mbox{\scriptsize{gw}}}^{2}(\nu)}{\nu^{6}S_{h}^{1}(\nu)S_{h}^{2}(\nu)}d\nu\right]
\label{SN}
\end{equation}
where $S_{h}^{1}$ and $S_{h}^{2}$ are the spectral noise densities, $T$ is the integration time (we are considering $T=1$yr), $\gamma(\nu)$ is the overlap reduction function, which depends on the relative positions and orientations of the two interferometers; and $\Omega_{\mbox{\scriptsize{gw}}}$ is the energy density parameter, which is given by\cite{araujo}
\begin{equation}
\Omega_{\mbox{\scriptsize{gw}}}=\frac{4\pi^{2}}{3H_{0}^{2}}\nu^{2}h_{\mbox{\scriptsize{BG}}}^{2}
\label{omega}
\end{equation}

In Table \ref{SNR} we show the results for S$/$N for the three families of compact binaries. For pairs of ILIGO, we have S$/$N$<1$ for the three families of compact binaries, which indicates that there is no possibility of detection; for pairs of ELIGO and ALIGO we have S$/$N$>1$ for the BHNS and DNS systems, where BHNS systems have the highest possibility of detection; for pairs of ET, the S$/$N is higher than $\sim 10^{3}$ for the three families of systems.

\begin{table}
\caption{The S$/$N for pairs of ILIGO, ELIGO, ALIGO and ET}
\label{SNR}
\begin{tabular}{lllll}
\hline\noalign{\smallskip}
System & &(S$/$N)& & \\ &ILIGO & ELIGO & ALIGO & ET\\
\noalign{\smallskip}\hline\noalign{\smallskip}
DNS &$3.1\times 10^{-2}$ &$2.7  $ &$12.0 $ & $1.6\times 10^{4}$\\
BHNS &$5.6\times 10^{-2}$ &$4.2  $ &$16.0 $ & $3.1\times 10^{4}$\\
BBH &$2.0\times 10^{-3}$ &$0.18 $ &$0.69$ & $1.9\times 10^{3}$\\
\noalign{\smallskip}\hline
\end{tabular}
\end{table}

\section{Conclusions}
\label{conclusion}

Our main aim here was to develop an alternative method to calculate the background generated by cosmological compact binary systems during their periodic or quasiperiodic phases. We use here an analogy with a problem of statistical mechanics in order to perform such a calculation as well as taking into account the temporal variation of the orbital parameters of the systems. Such a background is of particular interest since it could well form an important foreground for the planned gravitational wave interferometers DECIGO, BBO, LISA or eLISA, ALIGO and ET.

This new tool for the calculation of the background has the advantage of being simple and versatile, because the distribution functions and other functions and parameters could be easily changed without the need of modifying the formalism.

Our results show that the background generated by the cosmological compact binaries during their periodic and quasiperiodic phases does not form a foreground for LISA, eLISA, BBO, DECIGO and ALIGO.  For the ET a foreground (confusion noise) could marginally be formed. It is worth recalling that in the present study, we only considered systems in circular orbits, and the inclusion of the eccentricity will be investigated in a future study. Besides, we will use the formalism developed here in the calculation of the background generated by the coalescence compact binary systems.

We also investigated if the backgrounds studied here could be detected using a cross-correlation of two interferometric detectors. We note that pairs of ALIGOs and ETs could in principle detect such backgrounds.

Given that by the time ET becomes operative, there will probably be some advanced configuration of VIRGO, and then, it will be possible
to correlate them. Such a cross correlation would be, in principle, much more sensitive than a pair of ALIGOs. Therefore, it
will be possible to investigate the kind of backgrounds studied here.

\begin{acknowledgements}
EFDE would like to thank the Brazilian agencies CAPES and FAPESP for their financial support and the Astrophysics Division of National Institute for Space Research for providing the structure necessary for this research work. JCNA would like to thank FAPESP and CNPq for the partial financial support.
\end{acknowledgements}

\appendix

\section{Appendix}
\label{appenA}

Using the definition given by (\ref{dist1}), we have
\begin{equation}
dn=H(\nu)d\nu ,
\end{equation}
which is the fraction of systems originated at time $t_{0}$ with orbital frequencies in the interval $d\nu$.
Using the binary formation rate corresponding to each type of system, we have
\begin{equation}
\frac{dn}{d\nu dVdt_{0}}=n_{\mbox{\scriptsize{bin}}}(t_{0})H(\nu),
\end{equation}
where the subindex $bin$ refers to each type of system (DNS, BHNS and BBH).
Now, integrating over $dt_{0}$, we get
\begin{equation}
\frac{dn}{dV}=\left[\int n_{\mbox{\scriptsize{bin}}}(t_{0})H(\nu)dt_{0}\right]d\nu ,
\label{app1}
\end{equation}
where the expression in brackets is the number of systems per unit frequency interval and per comoving volume, which may be used as a distribution function in the calculation of the background.

\end{document}